\documentclass[prd,twocolumn,floatfix,amsmath,nofootinbib,amssymb,floatfix]{revtex4}
\usepackage{graphicx,color,dcolumn,booktabs,bm}
\usepackage{longtable,lscape}
\usepackage{pdfpages}
\usepackage{txfonts}
\usepackage{overpic}
\usepackage{amssymb}
\usepackage{makecell}
\usepackage{indentfirst}
\usepackage{feynmf}   
\usepackage{slashed}  
\usepackage{cases}
\usepackage{color}
\usepackage{multirow}
\usepackage{threeparttable}
\usepackage{epstopdf}
\usepackage{enumerate}
\usepackage{diagbox}
\usepackage{graphicx,color,dcolumn,booktabs,bm}
\usepackage[colorlinks,
            citecolor=blue,
            anchorcolor=red,
            menucolor=red,
            linkcolor=red,
            filecolor=red,
            runcolor=red,
            urlcolor=blue,
            frenchlinks=red]{hyperref}
\begin{document}
\title{The newly observed $\Omega_c(3327)$: A good candidate for a $D$-wave charmed baryon}
\author{Si-Qiang Luo$^{1,2,3,4,5}$}\email{luosq15@lzu.edu.cn}
\author{Xiang Liu$^{1,3,4,5}$}\email{xiangliu@lzu.edu.cn}
\affiliation{
$^1$School of Physical Science and Technology, Lanzhou University, Lanzhou 730000, China\\
$^2$School of mathematics and statistics, Lanzhou University, Lanzhou 730000, China\\
$^3$Research Center for Hadron and CSR Physics, Lanzhou University and Institute of Modern Physics of CAS, Lanzhou 730000, China\\
$^4$Lanzhou Center for Theoretical Physics, Key Laboratory of Theoretical Physics of Gansu Province, and Frontiers Science Center for Rare Isotopes, Lanzhou University, Lanzhou 730000, China\\
$^5$Key Laboratory of Quantum Theory and Applications of MoE, Lanzhou University, Lanzhou 730000, China}

\begin{abstract}
The newly observed $\Omega_c(3327)$
gives us a good chance to construct the $\Omega_c$ charmed baryon family. In this work, we carry out the mass spectrum analysis by a non-relativistic potential model using Gaussian Expansion Method, and the study of its two-body Okubo-Zweig-Iizuka allowed strong decay behavior. Our results imply that the $\Omega_c(3327)$ is good candidate of $\Omega_c(1D)$ state with $J^P=5/2^+$. We also predict the spectroscopy behavior of other $\Omega_c(1D)$ states, which may provide further clues to their search.
\end{abstract}
\maketitle


Very recently, the LHCb Collaboration observed two new hadronic states, $\Omega_c(3185)$ and $\Omega_c(3327)$, in the $\Xi_c^+K^-$ invarant spectrum~\cite{LHCb:2023rtu}. Their resonance parameters are given by
\begin{equation}
\begin{split}
M_{\Omega_c(3185)}=&3185.1\pm1.7^{+7.4}_{-0.9}\pm0.2\,{\rm MeV},\\
\Gamma_{\Omega_c(3185)}=&50\pm7^{+10}_{-20}\,{\rm MeV},\\
M_{\Omega_c(3327)}=&3327.1\pm1.2^{+0.1}_{-1.3}\pm0.2\,{\rm MeV},\\
\Gamma_{\Omega_c(3327)}=&20\pm5^{+13}_{-1}\,{\rm MeV}.\\
\end{split}
\end{equation}
In this experimental analysis~\cite{LHCb:2023rtu}, LHCb also confirmed the existence of the five narrow $\Omega_c$ states including the $\Omega_c(3000)$, $\Omega_c(3050)$, $\Omega_c(3065)$, $\Omega_c(3090)$, and $\Omega_c(3119)$, which were first reported in 2017~\cite{LHCb:2017uwr}. Also,
the $\Omega_c(3000)$, $\Omega_c(3050)$, $\Omega_c(3065)$, and $\Omega_c(3090)$ was also confirmed by the Belle Collaboration via the $e^+e^-$ collision~\cite{Belle:2017ext}, and by LHCb in
the $\Omega_b^-\to\Omega_c^0(X)\pi^-\to \Xi_c^+K^-\pi^-$ process~\cite{LHCb:2021ptx}. Obviously, associated with these excited states with the two $1S$ states $\Omega_c(2695)$ and $\Omega_c(2700)$, these new observations of the $\Omega_c$ states give us a good chance to construct the $\Omega_c$ charmed baryon family.

There have been some theoretical studies focusing on the $\Omega_c(3000)$, $\Omega_c(3050)$, $\Omega_c(3065)$, $\Omega_c(3090)$, and $\Omega_c(3119)$, which suggest that these five $\Omega_c$ states are good candidates of $\Omega_c(1P)$ or $\Omega_c(2S)$~\cite{Agaev:2017jyt,Chen:2017sci,Karliner:2017kfm,Wang:2017hej,Wang:2017vnc,Padmanath:2017lng,Cheng:2017ove,Wang:2017zjw,Zhao:2017fov,Chen:2017gnu,Agaev:2017lip,Yang:2020zrh,Yang:2017qan,Ali:2017wsf}. For the newly observed $\Omega_c(3185)$, its resonance parameter is close to that of the $\Omega_c(3188)$  discovered by LHCb and Belle \cite{LHCb:2017uwr,Belle:2017ext}. In principle, we can treat the $\Omega_c(3185)$ and the $\Omega_c(3188)$ in the same procedures. These efforts have made great progress in constructing the $\Omega_c$ charmed baryon family, where there are suitable candidates for the $1S$, $1P$, and $2S$ states of $\Omega_c$ (see review articles \cite{Cheng:2015iom,Chen:2016spr,Cheng:2021qpd,Chen:2022asf}). However, our knowledge of higher states in the $\Omega_c$ family is still lacking. Further efforts should therefore be made.

Against this research background, the newly observed $\Omega_c(3327)$ is timely, since deciphering the nature of the $\Omega_c(3327)$ may provide useful clues to establish the $1D$ states of the $\Omega_c$ family, which will be a major task of this work.

To achieve this goal, we perform the mass spectrum analysis and the study of its two-body Okubo-Zweig-Iizuka (OZI)-allowed strong decay behavior. When performing the mass spectrum analysis, we adopt a non-relativistic potential model~\cite{Luo:2019qkm,Luo:2021dvj} with the help of the Gaussian expansion method (GEM)~\cite{Hiyama:2003cu}, which can promote the accuracy of the calculation. Obtaining the mass spectrum of the $\Omega_c$ family is only one aspect of deciphering the property of the $\Omega_c(3327)$. In the following, we should focus on its two-body OZI-allowed strong decay behavior. The numerical spatial wave function obtained by the mass spectrum analysis can be used as input, by which we can avoid the uncertainty of simply applying the Simple Harmonic Oscillator (SHO) wave function to deal with the spatial wave function. In the concrete calculation of its strong decay behavior, the quark pair creation  model~\cite{Micu:1968mk} is adopted, which is an effective approach to estimate the partial widths of the strong decay of baryon. Later, we will briefly present some details of the QPC model. In general, we can finally suggest that the newly observed $\Omega_c(3327)$ is a $1D$ state of $\Omega_c$ with spin parity quantum number $J^P=5/2^+$. Taking this opportunity, we also give some further predictions of its strong decay mode, which can be used to test this assignment. In fact, there are six $1D$ states of $\Omega_c$. In this work, their masses and two-body OZI-allowed decay behaviors are predicted, which is useful for future searches for them. 



\begin{table}[htbp]
\caption{The parameters involved in the adopted potential model.}
\label{tab:parameter}
\renewcommand\arraystretch{1.25}
\begin{tabular*}{86mm}{@{\extracolsep{\fill}}m{10mm}m{15mm}<{\centering}m{15mm}<{\centering}m{15mm}<{\centering}m{15mm}<{\centering}}
\toprule[1.00pt]
\toprule[1.00pt]
system               &$\alpha_s$ &$b$ (GeV$^2$) &$\sigma$ (GeV) &$C$ (GeV)\\
\midrule[0.75pt]
$\Xi_c/\Xi_c^\prime$ &0.548      &0.144         &1.732          &-0.711   \\
$\Omega_c$           &0.578      &0.144         &1.732          &-0.688   \\
meson                &0.578      &0.144         &1.020          &-0.685   \\
\midrule[0.75pt]
\multicolumn{5}{c}{\mbox{$m_{u/d}=0.370~{\rm GeV}~m_{s}=0.600~{\rm GeV}~m_{c}=1.880~{\rm GeV}$}}\\
\bottomrule[1.00pt]
\bottomrule[1.00pt]
\end{tabular*}
\end{table}

As a first step, we need the mass spectrum of the singly heavy baryons. Here, a non-relativistic potential model has been adopted. The corresponding Hamiltonian is \cite{Yoshida:2015tia,Luo:2019qkm,Luo:2021dvj}
\begin{equation}\label{eq:H}
\hat{H}=\sum\limits_{i}\left(m_i+\frac{p_i^2}{2m_i}\right)+\sum\limits_{i<j}V_{ij},
\end{equation}
where $m_i$ and $p_i$ are the mass and momentum of the $i$-th constituent quark, respectively. The last term $V_{ij}$ in Eq.~(\ref{eq:H}) denotes the quark-quark interaction, which is expressed by
\begin{equation}\label{eq:Vij}
V_{ij}=H_{ij}^{\rm conf}+H_{ij}^{\rm hyp}+H_{ij}^{\rm so(cm)}+H_{ij}^{\rm so(tp)}.
\end{equation}
In Eq.~(\ref{eq:Vij}), the first term is spin-independent, i.e.,
\begin{equation}\label{eq:Hconf}
H_{ij}^{\rm conf}=-\frac{2}{3}\frac{\alpha_s}{r_{ij}}+\frac{b}{2}r_{ij}+\frac{1}{2}C.
\end{equation}
Here, the $\alpha_s$, $b$, and $C$ are the coupling constant of the one-gluon-exchange (OGE), the strength of the linear confinement, and the renormalized  mass constant, respectively. The remaining parts of Eq.~(\ref{eq:Vij}) are spin-dependent potentials, including the hyperfine interaction $H_{ij}^{\rm hyp}$, the color-magnetic term $H_{ij}^{{\rm so(cm)}}$, and the Thomas-precession piece $H_{ij}^{{\rm so(tp)}}$, which are written as
\begin{equation}\label{eq:Hyp}
H_{ij}^{\rm hyp}=\frac{2\alpha_s}{3m_im_j}\left[\frac{8\pi}{3}\tilde{\delta}(r_{ij}){\bf s}_i\cdot{\bf s}_j+\frac{1}{r_{ij}^3}S({\bf r},{\bf s}_i,{\bf s}_j)\right],
\end{equation}
\begin{equation}\label{eq:Hsocm}
\begin{split}
H_{ij}^{{\rm so(cm)}}=&\frac{2\alpha_s}{3r_{ij}^3}\left(\frac{{\bf r}_{ij}\times{\bf p}_i\cdot{\bf s}_i}{m_i^2}-\frac{{\bf r}_{ij}\times{\bf p}_j\cdot{\bf s}_j}{m_j^2}\right.\\
&\left.-\frac{{\bf r}_{ij}\times{\bf p}_j\cdot{\bf s}_i-{\bf r}_{ij}\times{\bf p}_i\cdot{\bf s}_j}{m_im_j}\right),
\end{split}
\end{equation}
and
\begin{equation}\label{eq:Hsotp}
H_{ij}^{{\rm so(tp)}}=-\frac{1}{2r_{ij}}\frac{\partial H_{ij}^{\rm conf}}{\partial r_{ij}}\left(\frac{{\bf r}_{ij}\times{\bf p}_i\cdot{\bf s}_i}{m_i^2}-\frac{{\bf r}_{ij}\times{\bf p}_j\cdot{\bf s}_j}{m_j^2}\right).
\end{equation}
The Gaussian smearing function $\tilde{\delta}(r_{ij})$ and tensor operator $S({\bf r},{\bf s}_i,{\bf s}_j)$ in Eq. (\ref{eq:Hyp}) are defined as
\begin{equation}
\tilde{\delta}(r)=\frac{\sigma^3}{\pi^{3/2}}{\rm e}^{-\sigma^2r^2},~~~S({\bf r},{\bf s}_i,{\bf s}_j)=\frac{3{\bf s}_i\cdot{\bf r}_{ij}{\bf s}_j\cdot{\bf r}_{ij}}{r_{ij}^2}-{\bf s}_i\cdot{\bf s}_j,
\end{equation}
respectively. For the quark-antiquark potential $V_{ij}^{q\bar{q}}$ in the meson system, we simply take $V_{ij}^{q\bar{q}}=2V_{ij}^{qq}$, since the color factor of the quark-(anti)quark within a meson is exactly twice that of the baryon system.

\begin{figure}
\begin{center}
\includegraphics[width=8.6cm,keepaspectratio]{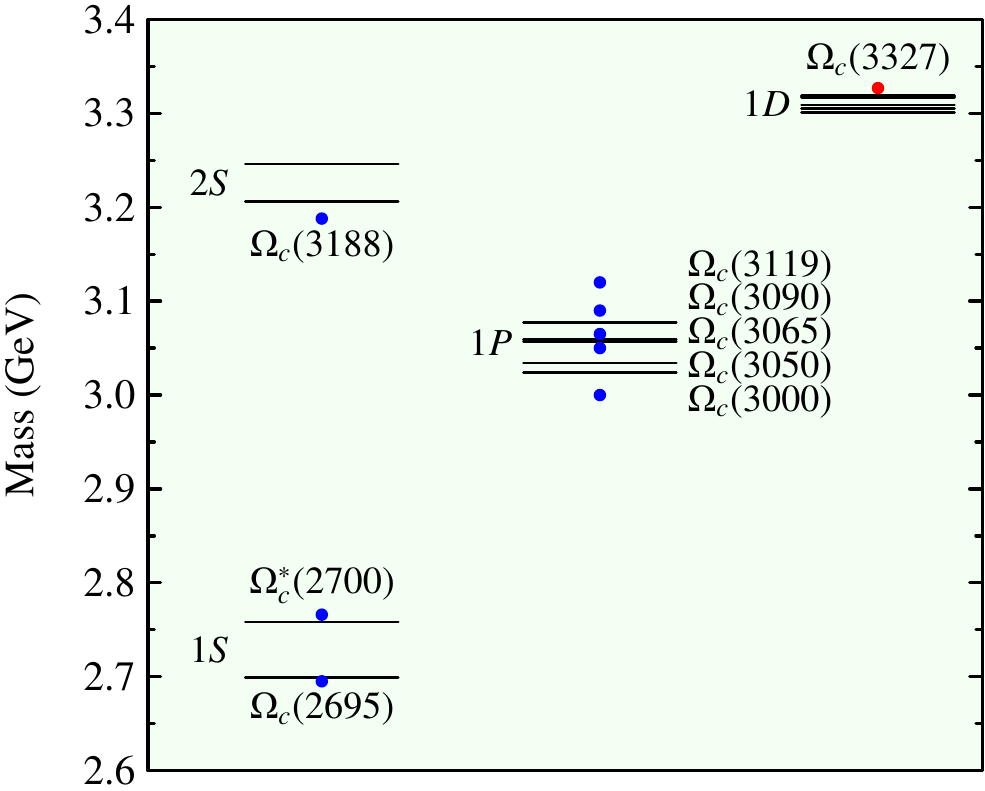}
\caption{(Color online.) The calculated mass spectrum of the $\Omega_c$ and the comparison with the experimental data. The short lines represent the calculated results, while the blue points are obtained from the experimental data taken from the Particle Data Group (PDG)~\cite{ParticleDataGroup:2020ssz}. The newly observed $\Omega_c(3327)$~\cite{LHCb:2023rtu} is marked by the red point.}
\label{fig:spectrum}
\end{center}
\end{figure}

\begin{table}
\caption{The comparisons of our calculated mass spectrum of $\Omega_c$ charmed baryon with the results from other theoretical groups.}
\label{tab:Omegacspectrum}
\renewcommand\arraystretch{1.25}
\begin{tabular*}{86mm}{@{\extracolsep{\fill}}lccccc}
\toprule[1.00pt]
\toprule[1.00pt]
States                   &PDG~\cite{ParticleDataGroup:2020ssz} &This Work &Ref.~\cite{Ebert:2011kk} &Ref.~\cite{Yu:2022ymb}&Ref.~\cite{Yoshida:2015tia}\\
\midrule[0.75pt]
$\Omega_{c}(1S,1/2^+)$   &2695                                 &2699      &2698                     &2699 &2731           \\
$\Omega_{c}(2S,1/2^+)$   &                                     &3206      &3088                     &3150 &3227           \\
$\Omega_{c}(1S,3/2^+)$   &2766                                 &2758      &2768                     &2762 &2779           \\
$\Omega_{c}(2S,3/2^+)$   &                                     &3246      &3123                     &3197 &3257           \\
$\Omega_{c0}(1P,1/2^-)$  &                                     &3034      &2966                     &3057 &3030           \\
$\Omega_{c1}(1P,1/2^-)$  &                                     &3024      &3055                     &3045 &3048           \\
$\Omega_{c1}(1P,3/2^-)$  &                                     &3059      &3029                     &3062 &3033           \\
$\Omega_{c2}(1P,3/2^-)$  &                                     &3057      &3054                     &3039 &3056           \\
$\Omega_{c2}(1P,5/2^-)$  &                                     &3077      &3051                     &3067 &3057           \\
$\Omega_{c1}(1D,1/2^+)$  &                                     &3301      &3287                     &3304 &3292           \\
$\Omega_{c1}(1D,3/2^+)$  &                                     &3305      &3282                     &3313 &3285           \\
$\Omega_{c2}(1D,3/2^+)$  &                                     &3318      &3298                     &3304 &$\cdots$       \\
$\Omega_{c2}(1D,5/2^+)$  &                                     &3319      &3286                     &3314 &3288           \\
$\Omega_{c3}(1D,5/2^+)$  &                                     &3309      &3297                     &3304 &3299           \\
$\Omega_{c3}(1D,7/2^+)$  &                                     &3317      &3283                     &3315 &$\cdots$       \\
\bottomrule[1.00pt]
\bottomrule[1.00pt]
\end{tabular*}
\end{table}

It is convenient to use the $\rho$-mode and $\lambda$-mode to distinguish the different excited modes of singly charmed baryon\footnote{The $\rho$-mode denote the excitation between two light quarks ($q_1$ and $q_2$), while the $\lambda$-mode represents the excitation between the light quark cluster and heavy quark ($Q$).}. The following basis
\begin{equation}\label{eq:basisJM}
|JM\rangle=|[[s_{q_1}s_{q_2}]_{s_\ell}[n_\rho n_\lambda l_\rho l_\lambda]_L]_{j_\ell}s_Q]_{JM}\rangle,
\end{equation}
is used to calculate the masses of singly charmed baryons. Here, $s_{q_1}$ and $s_{q_2}$ are the spins of the light flavor quarks, while $s_{Q}$ is the spin of the heavy flavor quark. The $s_{\ell}$ in Eq.~(\ref{eq:basisJM}) denotes the total spin of the two light flavor quarks. $n_{\rho/\lambda}$ and $l_{\rho/\lambda}$ are the radial and orbital quantum numbers, respectively. $L$ is the total orbital angular momentum of the system. And $j_\ell$ stands for the total angular momentum of the light degree-of-freedom.

In this work, we employ the GEM~\cite{Hiyama:2003cu} to solve the Schr\"odinger equations of the mesons and singly charmed baryons. In our calculation, the values of the $\alpha_s$, $b$, $\sigma$, $C$, and the consistent quark masses in the quark potential model were constrained by the well-established mesons and charm baryons. The concrete values of the parameters are listed in Table~\ref{tab:parameter}.

The numerical results of the mass spectrum of singly charmed baryon are shown in Fig.~\ref{fig:spectrum}. So far, all observed singly charmed baryons belong to $\lambda$-mode excited state~\cite{Chen:2016iyi,Chen:2021eyk}, which is the reason why in this work we mainly focus on the assignment of the $\lambda$-mode exited state to the $\Omega_c(3327)$. Thus, six $\lambda$-mode $1D$ $\Omega_c$ states are $\Omega_{c1}(1D,1/2^+)$, $\Omega_{c1}(1D,3/2^+)$, $\Omega_{c2}(1D,3/2^+)$, $\Omega_{c2}(1D,5/2^+)$, $\Omega_{c3}(1D,5/2^+)$ and $\Omega_{c3}(1D,7/2^+)$, where the subscripts $1,2,3,$ denote the $j_\ell$ quantum number of the corresponding states. The same notation is also applied to denote other $\Omega_c$ states in Table~\ref{tab:Omegacspectrum}. Here, most of the calculated results are in good agreement with the experimental data. In this way, we test the feasibility of the adopted potential model used to represent the mass spectrum of singly charmed baryon. Furthermore, in order to refer to the details of the $\Omega_c$ mass spectrum, we also collect our results and make comparison with those of different theoretical groups in Table~\ref{tab:Omegacspectrum}. And similar results have also been obtained in Refs.~\cite{Ebert:2011kk,Yu:2022ymb,Yoshida:2015tia,Roberts:2007ni,Yamaguchi:2014era,Shah:2016nxi,Garcia-Tecocoatzi:2022zrf,Mao:2017wbz}. We find that the mass of the newly observed $\Omega_c(3327)$ is close to the calculated mass range of the $\Omega_c(1D)$ states, suggesting that the $\Omega_c(3327)$ can be as a good $\Omega_c(1D)$ candidate. However, since the mass splits of the six $\Omega_c(1D)$ are too small, it is difficult to determine the spin-parity quantum number of the $\Omega_c(3327)$ directly from this mass spectrum analysis. We need to further unravel its nature by investigating its two-body OZI-allowed strong decay behaviors.

\begin{table}[htbp]
\caption{The $\beta$ values (in units of GeV) of these involved mesons and singly charmed baryons.}
\label{tab:beta}
\renewcommand\arraystretch{1.25}
\begin{tabular*}{86mm}{@{\extracolsep{\fill}}lcclcclcc}
\toprule[1.00pt]
\toprule[1.00pt]
States                   &$\beta_\rho$ &$\beta_\lambda$   &States                     &$\beta_\rho$ &$\beta_\lambda$   &States                   &$\beta_\rho$   &$\beta_\lambda$ \\
\midrule[0.75pt]
$\Xi_c(1S)$              &0.304        &0.384             &$\Xi^\prime_c(1S)$         &0.252        &0.384             &$\Omega_c(1S)$           &0.288          &0.420           \\
$\cdots$                 &$\cdots$     &$\cdots$          &$\Xi^*_c(1S)$              &0.243        &0.358             &$\Omega_c^*(1S)$         &0.275          &0.389           \\
$\Xi_c(2S)$              &0.262        &0.207             &$\Xi^\prime_c(2S)$         &0.214        &0.210             &$\Omega_c(2S)$           &0.230          &0.231           \\
$\cdots$                 &$\cdots$     &$\cdots$          &$\Xi^*_c(2S)$              &0.217        &0.201             &$\Omega_c^*(2S)$         &0.236          &0.218           \\
$\Xi_c(1P)$              &0.278        &0.258             &$\Xi^\prime_c(1P)$         &0.229        &0.258             &$\Omega_c(1P)$           &0.257          &0.279           \\
$\Xi_c(1D)$              &0.267        &0.201             &$\Xi^\prime_c(1D)$         &0.219        &0.199             &$\Omega_c(1D)$           &0.244          &0.212           \\
\Xcline{4-9}{0.75pt}
$\Xi$                    &0.287        &0.317             &\multicolumn{6}{c}{$\beta_\pi=0.409$~~$\beta_{s\bar{s}(1^1S_0)}=0.402$~~$\beta_K=0.385$}                                 \\
$\Xi^*$                  &0.258        &0.265             &\multicolumn{6}{c}{$\beta_D=0.357$~~$\beta_{D^*}=0.307$}                                                               \\
\bottomrule[1.00pt]
\bottomrule[1.00pt]
\end{tabular*}
\end{table}

\begin{table*}
\centering
\caption{The partial and total decay widths of the $\lambda$-mode excited $\Omega_c(1D)$ in units of MeV. The forbidden couplings are denoted by the symbol ``$\times$''. 
The number ``0.0" implies that the partial decay width is less than 0.1 MeV. Here, We do not consider the mixture between $\Omega_{c1}(1D,3/2^+)$ and $\Omega_{c3}(1D,3/2^+)$ and the mixture between $\Omega_{c2}(1D,5/2^+)$ and $\Omega_{c3}(1D,5/2^+)$. When presenting these decay behaviors, the masses of these six $1D$ states of $\Omega_c$ are from the experimental mass of $\Omega_c(3327)$.}
\label{tab:Omegac1D}
\renewcommand\arraystretch{1.15}
\begin{tabular*}{178mm}{@{\extracolsep{\fill}}lcccccc}\toprule[1.00pt]
\toprule[1.00pt]
Decay channels
&$\Omega_{c1}(1D,1/2^+)$&$\Omega_{c1}(1D,3/2^+)$&$\Omega_{c2}(1D,3/2^+)$&$\Omega_{c2}(1D,5/2^+)$&$\Omega_{c3}(1D,5/2^+)$&$\Omega_{c3}(1D,7/2^+)$\\
\midrule[0.75pt]
$\Xi_c(2470)\bar{K}$&2.7&2.7&$\times$&$\times$&13.4&13.4\\
$\Xi_c(2790)\bar{K}$&125.0&0.5&1.1&0.4&3.6&0.0\\
$\Xi_c(2815)\bar{K}$&0.0&114.1&0.0&0.1&0.0&0.3\\
$\Xi_c^\prime(2580)\bar{K}$&3.9&0.9&8.7&2.6&3.0&1.7\\
$\Xi_c^*(2645)\bar{K}$&2.7&6.7&5.2&15.8&2.2&3.0\\
$\Omega_c(2695)\eta$&0.4&0.1&1.0&0.0&0.0&0.0\\
$\Omega_c(2765)\eta$&0.0&0.0&0.0&0.1&0.0&0.0\\
$\Xi D$&244.9&15.3&137.8&31.3&2.2&80.6\\
$\Xi D^*$&5.6&16.3&3.8&10.2&0.0&0.0\\
\midrule[0.75pt]
Total&385.2&156.6&157.6&60.5&24.4&99.0\\
Exp. &    &      &     &    &$20\pm5^{+13}_{-1}$~\cite{LHCb:2023rtu} &\\
\bottomrule[1.00pt]
\bottomrule[1.00pt]
\end{tabular*}
\end{table*}


In this work, the quark pair creation (QPC) model~\cite{Micu:1968mk,LeYaouanc:1972vsx,LeYaouanc:1973ldf,Barnes:2007xu,Ackleh:1996yt} is employed to calculate the strong decays of the $\Omega_c(1D)$ states. The QPC model has been widely used to study the strong decays of the singly charmed baryons~\cite{Chen:2007xf,Chen:2016iyi,Chen:2017aqm,Chen:2017gnu,Chen:2018orb,Chen:2018vuc,Chen:2019ywy,Lu:2018utx,Liang:2019aag,Lu:2019rtg,Lu:2020ivo,Liang:2020kvn,Xiao:2017dly,Zhao:2016qmh,Zhao:2017fov,Ye:2017yvl,Ye:2017dra,Zhao:2020tpf,Guo:2019ytq,Yang:2018lzg}. The transition operator of the QPC model is
\begin{equation}\label{eq:toperator}
\begin{split}
\hat{\cal T}=&-3\gamma\sum_{m}\langle 1,m;1,-m|0,0\rangle\int{\rm d}^3{\bf p}_i\;{\rm d}^3{\bf p}_j\;\delta({\bf p}_i+{\bf p}_j)\\
          &\times\mathcal{Y}_1^m\left(\frac{{\bf p}_i-{\bf p}_j}{2}\right)\omega_0^{(i,j)}\phi_0^{(i,j)}\chi_{1,-m}^{(i,j)}b^\dagger_i({\bf p}_i)d^\dagger_j({\bf p}_j),
\end{split}
\end{equation}
where $\omega$, $\phi$, $\chi$ and $\mathcal{Y}$ are the color, flavour, spin, and spatial functions of the quark pair, respectively. The $b_{i}^\dagger$ and $d_{j}^\dagger$ are the quark and anti-quark creation operators, respectively. The dimensionless parameter $\gamma$ describes the strength of a quark-antiquark pair created from the vacuum, which is fixed as $\gamma=9.58$ by fitting the experimental width of the $\Sigma_c^*(2520)$ state \cite{ParticleDataGroup:2020ssz}. The partial wave amplitude for a decay process $A\to BC$ with relative spin $S_{BC}$ and orbital angular momentum $L_{BC}$ between $BC$ could be expressed by
\begin{equation}
{\cal M}_{A\to BC}^{L_{BC}S_{BC}}(p)=\langle BC,L_{BC},S_{BC},p|\hat{\cal T}|A\rangle,
\end{equation}
where $p$ is the momentum of the final state $B$. The partial width could be calculated by
\begin{equation}
\Gamma_{A\to BC}^{L_{BC}S_{BC}}=2\pi\frac{\sqrt{M_B^2+p^2}\sqrt{M_C^2+p^2}}{M_A}p|{\cal M}_{A\to BC}^{L_{BC}S_{BC}}(p)|^2.
\end{equation}

\begin{figure}
\begin{center}
\includegraphics[width=8.6cm,keepaspectratio]{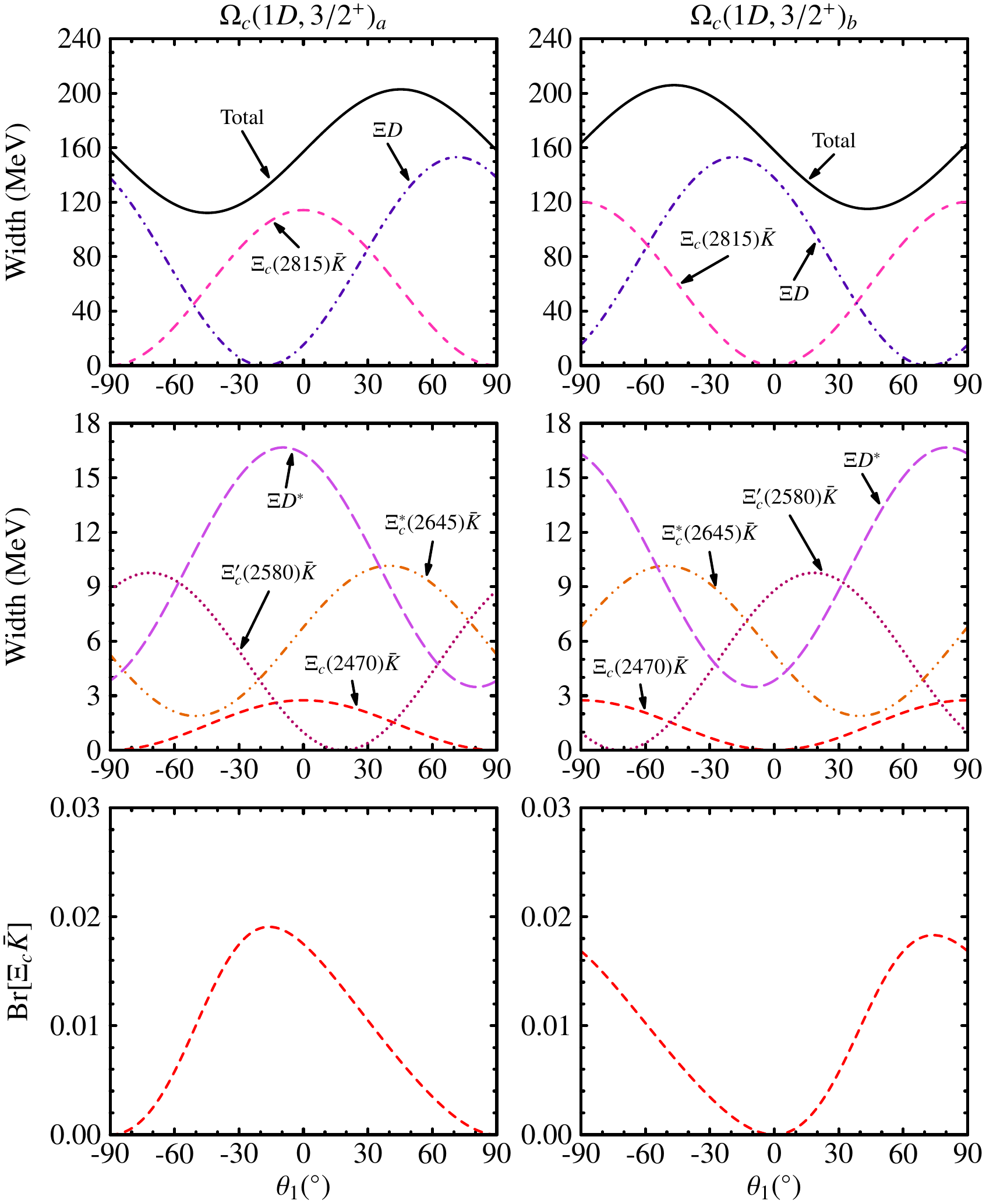}
\caption{(Color online.) The decay widths and branch ratios dependence on the mixing angles for $\Omega_c(1D,3/2^+)_a$ and $\Omega_c(1D,3/2^+)_b$. The masses are taken from measured mass of $\Omega_c(3327)$~\cite{LHCb:2023rtu}. Some small widths are not showing here but counted into the total widths.}
\label{fig:Omegac1D_1}
\end{center}
\end{figure}

\begin{figure}
\begin{center}
\includegraphics[width=8.6cm,keepaspectratio]{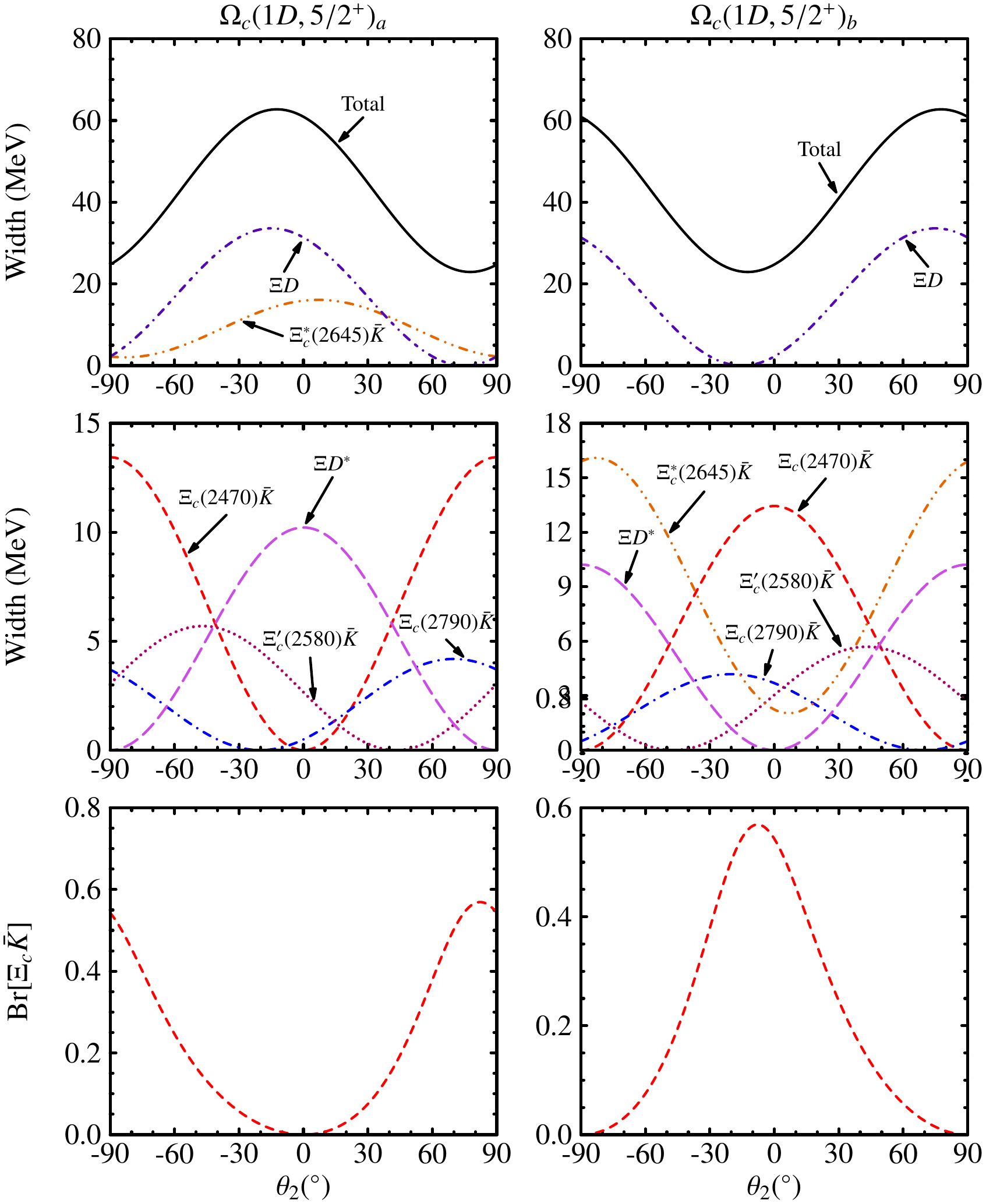}
\caption{(Color online.) The decay widths and branch ratios dependence on the mixing angles for $\Omega_c(1D,5/2^+)_a$ and $\Omega_c(1D,5/2^+)_b$. The conventions are the consistent with the Table~\ref{fig:Omegac1D_1}.}
\label{fig:Omegac1D_2}
\end{center}
\end{figure}

In the calculations of meson decays~\cite{Close:2005se,Song:2015nia,Barnes:2007xu,Ackleh:1996yt}, it is convenient to employ the simple harmonic oscillator (SHO) wave functions to depict the spatial structures of hadrons, i.e.,
\begin{equation}\label{eq:sho}
\begin{split}
R^p_{nlm}(\beta,{\bf P})=&\frac{(-1)^n(-{\mathrm i})^l}{\beta^{\frac{3}{2}+l}}\sqrt{\frac{2n!}{\Gamma(n+l+\frac{3}{2})}}L_{n}^{l+\frac{1}{2}}({P^2}/{\beta^2})\\
&\times {\mathrm e}^{-\frac{P^2}{2\beta^2}}P^l Y_{l m}(\Omega_{\bf P}),
\end{split}
\end{equation}
where $n$, $l$, and $m$ represent the radial, orbital, and magnetic quantum numbers, respectively. The $\beta$ value in Eq.~(\ref{eq:sho}) is a parameter for scaling the SHO wave function, which could be calculated from the GEM. By solving the Schr\"odinger equations with the GEM, one could obtain the spatial wave function $\phi^r_{n_\rho n_\lambda l_\rho l_\lambda m_\rho m_\lambda}({\bm \rho},{\bm \lambda})$, which is the summation of a series of Gaussian wave functions. Since a singly charmed baryon contains two spatial degrees-of-freedom, two $\beta$ values ($\beta_\rho$ and $\beta_\lambda$) are introduced in the SHO wave functions to reduce $\phi^r_{n_\rho n_\lambda l_\rho l_\lambda m_\rho m_\lambda}({\bm \rho},{\bm \lambda})$, which can be extracted with
\begin{equation}
\begin{split}
\frac{1}{\beta_\rho^2}=&\int|\phi^r_{n_\rho n_\lambda l_\rho l_\lambda m_\rho m_\lambda}({\bm \rho},{\bm \lambda})|{\bm \rho}^2{\rm d}^3{\bm \rho}{\rm d}^3{\bm \lambda},\\
\frac{1}{\beta_\lambda^2}=&\int|\phi^r_{n_\rho n_\lambda l_\rho l_\lambda m_\rho m_\lambda}({\bm \rho},{\bm \lambda})|{\bm \lambda}^2{\rm d}^3{\bm \rho}{\rm d}^3{\bm \lambda}.
\end{split}
\end{equation}
Here, the formula for calculating the $\beta$ value is slightly different from the Refs.~\cite{Close:2005se,Song:2015nia}, but the calculations for the decay widths of singly heavy baryons~\cite{Chen:2016iyi,Chen:2017aqm,Chen:2017gnu,Chen:2018orb,Chen:2018vuc,Chen:2019ywy} suggest that this scheme is also an effective approach. Then the spatial wave functions could be written by the following approximation
\begin{equation}
\phi^r_{n_\rho n_\lambda l_\rho l_\lambda m_\rho m_\lambda}({\bm \rho},{\bm \lambda})\approx R^r_{n_\rho l_\rho m_\rho}(\beta_\rho,{\bm \rho})R^r_{n_\lambda l_\lambda m_\lambda}(\beta_\lambda,{\bm \lambda}).
\end{equation}
Since it is convenient to calculated decay widths in the momentum space with the QPC model, we transform the wave function into the  momentum representation, i.e.,
\begin{equation}
\phi^p_{n_\rho n_\lambda l_\rho l_\lambda m_\rho m_\lambda}({\bf p}_\rho,{\bf p}_\lambda)\approx R^p_{n_\rho l_\rho m_\rho}(\beta_\rho,{\bf p}_\rho)R^p_{n_\lambda l_\lambda m_\lambda}(\beta_\lambda,{\bf p}_\lambda).
\end{equation}

There are six $1D$ states of $\Omega_c$ charmed baryon as shown in Fig. \ref{fig:spectrum}. The mass spectrum analysis only shows that the newly discovered $\Omega_c(3327)$ have relation to the $1D$ states of $\Omega_c$. The information of the total and partial decay widths of six $\Omega_c(1D)$ charmed baryons is crucial to decode the nature of the $\Omega_c(3327)$. In this subsection, we focus on the study of the partial decay widths of these discussed $\Omega_c(1D)$ charmed baryons.

In Table~\ref{tab:Omegac1D}, we list the calculated partial and total decay widths of six $1D$ states of $\Omega_c$, and make a comparison with the experimental data of the $\Omega_c(3327)$~\cite{LHCb:2023rtu}. We should note that the masses of the six $1D$ states of $\Omega_c$ are taken to be the mass of the $\Omega_c(3327)$ for ease to make comparison. We find that the measured width of the $\Omega_c(3327)$ is very close to the obtained total decay width of the $\Omega_{c3}(1D,5/2^+)$ state. In addition, the $\Xi_c(2470)\bar{K}$ channel has largest contribution to the total decay width, which can be reflected by the corresponding branching ratio
\begin{equation}
{\rm BR}[\Omega_{c3}(1D,5/2^+)\to \Xi_c(2470)\bar{K}]\approx55\%,
\end{equation}
which also explains why the $\Omega_c(3327)$ was first observed in its $\Xi_c^+(2470)K^-$ channel. Thus, according to this study, we may conclude that assigning the $\Omega_c(3327)$ as the $\Omega_{c3}(1D,5/2^+)$ state is suitable. If the $\Omega_c(3327)$ is the $\Omega_{c3}(1D,5/2^+)$ state, the partial widths of the $\Omega_c(3327)$ decays into $\Xi_c^\prime(2580)\bar{K}$, $\Xi D$, $\Xi_c^*(2645)\bar{K}$ and $\Xi_c(2790)\bar{K}$ are sizable.

We should introduce the decay behaviors of the other two $1D$ states of $\Omega_c$. $\Omega_{c1}(1D,1/2^+)$ is a very broad state, where $\Xi D$ and $\Xi_c(2790) \bar{K}$ have dominant contribution to the width of $\Omega_{c1}(1D,1/2^+)$. Although the total decay width of $\Omega_{c1}(1D,3/2^+)$ is close to that of $\Omega_{c2}(1D,3/2^+)$, their decay behaviors are different. Here, $\Xi_c(2815) \bar{K}$ and $\Xi D$ are the dominant decay modes of $\Omega_{c1}(1D,3/2^+)$ and $\Omega_{c2}(1D,3/2^+)$, respectively. The $\Omega_{c2}(1D,5/2^+)$ have a total width of 61.9 MeV, coming mainly from the $\Xi D$, $\Xi_c^*(2645)\bar{K}$ and $\Xi D^*$ channels. There exists $\Omega_{c3}(1D,7/2^+)$, where $\Xi D$ is the dominant contributor to the total decay width and  $\Xi_c(2470)\bar{K}$ is sizable. This obtained decay information is valuable for further experimental searches.

In reality, the mixing between $\Omega_{c2}(1D,3/2^+)$ and $\Omega_{c3}(1D,3/2^+)$ can happen, i.e.,
\begin{equation}\label{eq:1D32}
\left(\begin{array}{c}
|\Omega_c(1D,3/2^+)_a\rangle\\
|\Omega_c(1D,3/2^+)_b\rangle
\end{array}\right)
=\left(\begin{array}{cc}
 \cos \theta_1 & \sin \theta_1\\
-\sin \theta_1 & \cos \theta_1
\end{array}\right)
\left(\begin{array}{c}
|\Omega_{c2}(1D,3/2^+)\rangle\\
|\Omega_{c3}(1D,3/2^+)\rangle
\end{array}\right).
\end{equation}
Here, $\Omega_c(1D,3/2^+)_a$ and $\Omega_c(1D,3/2^+)_b$ are two physical states, where the subscripts `$a$' and `$b$' are applied to distinguish the physical states with lower and higher masses, respectively. $\theta_1$ is mixing angle, which can be estimated to be zero in the limit of the heavy quark mass $m_Q\to \infty$~\cite{Chen:2019ywy,Wang:2017vnc}. Note, however, that the charm quark mass is not heavy enough. Thus, the mixing angle $\theta_1$ is not zero in a realistic situation. We take a range of mixing angle  $-90^\circ\leq \theta_1\leq90^\circ$ to discuss the dependence of their total and partial decay widths on the mixing angle, which is shown in Fig.~\ref{fig:Omegac1D_1}.

There is also a mixture between $\Omega_{c2}(1D,5/2^+)$ and $\Omega_{c3}(1D,5/2^+)$, which is written as
\begin{equation}\label{eq:1D52}
\left(\begin{array}{c}
|\Omega_c(1D,5/2^+)_a\rangle\\
|\Omega_c(1D,5/2^+)_b\rangle
\end{array}\right)
=\left(\begin{array}{cc}
 \cos \theta_2 & \sin \theta_2\\
-\sin \theta_2 & \cos \theta_2
\end{array}\right)
\left(\begin{array}{c}
|\Omega_{c2}(1D,5/2^+)\rangle\\
|\Omega_{c3}(1D,5/2^+)\rangle
\end{array}\right).
\end{equation}
In this paper, we also show the dependence of their decay behavior on the mixing angle $\theta_2$ as given in Fig.~\ref{fig:Omegac1D_2}.
Since this mixing angle cannot be fixed by experimental data or theoretical input, we have to stop the discussion on the strong decay behaviors of these four mixing states.

In summary, stimulated by the observation of the $\Omega_c(3327)$ from LHCb \cite{LHCb:2023rtu}, in this work we decipher its nature by carrying out the analysis of the mass spectrum analysis and the calculation of the two-body OZI-allowed strong decay. Our result shows the possibility of assigning the $\Omega_c(3327)$ as a $1D$ state of $\Omega_c$, which has the spin parity quantum number $J^P=5/2^+$. We also notice a detail of its partial decay widths under this assignment, where $\Xi_c(2470)\bar{K}$ is its main decay channel in our calculation, which can naturally explain why the $\Omega_c(3327)$ was first observed in this decay channel. As by-product, we also predict the decay behavior of the other five $1D$ states of $\Omega_c$, which are still missing in the  experiment. Obviously, the present study can  provide some clues for their future exploration.

The study of this work opens a window for the construction of the $1D$ states of charmed baryon. With the accumulation of experimental data in LHCb and Belle II, we have a reason to believe that further experimental progress will be made. For the discussed $\Omega_c(3327)$, the measurement of its spin parity quantum number and the observation of its other decay modes are a crucial step to establish the $\Omega_c(3327)$ as the $1D$ states of charmed baryon, which will be a new task for experimentalists.

{\bf Note added}: When preparing the present manuscript, we noticed a recent paper on the the $\Omega_c(3327)$ in arXiv~\cite{Yu:2023bxn}, where the authors tried to explain the $\Omega_c(3327)$ as a $1D$ state of the $\Omega_c$ with $J^P=3/2^+$, which is different from our assignment to the $\Omega_c(3327)$. We also note an earlier paper \cite{Yao:2018jmc}. The decay widths of $\Omega(1D)$ was predicted by the chiral quark model, which supports to the $\Omega_c(3327)$ as a $\Omega(1D)$ state, which is consistent with our conclusion.

\section*{ACKNOWLEDGMENTS}

We would like to thank Bing Chen and Hai-Yang Cheng for useful discussions. This work is supported by the China National Funds for Distinguished Young Scientists under Grant No. 11825503, the National Key Research and Development Program of China under Contract No. 2020YFA0406400, the 111 Project under Grant No. B20063, the National Natural Science Foundation of China under Grant No. 12247101, the fundamental Research Funds for the Central Universities under Grant No. lzujbky-2022-sp02, and the project for top-notch innovative talents of Gansu province.

\end{document}